\documentclass[amsmath,amssymb,prx,twocolumn]{revtex4-2}

\usepackage{graphicx,dcolumn,txfonts}
\usepackage{verbatim}

\usepackage{sidecap}
\usepackage{hyperref}
\usepackage[utf8]{inputenc}
\usepackage{fourier} 
\usepackage{array}
\usepackage{makecell}
\usepackage{enumitem}

\begin{document}

\title{Flexible imitation suppresses epidemics through better vaccination}

\author{Soya Miyoshi}
\author{Marko Jusup}
\author{Petter Holme}
 \email{holme@cns.pi.titech.ac.jp}
\affiliation{Tokyo Tech World Research Hub Initiative (WRHI), Institute of Innovative Research, Tokyo Institute of Technology, Yokohama, Japan}

\begin{abstract}
\vspace{0.1cm}
\begin{center}
\noindent\textbf{ABSTRACT}\\    
\end{center}

The decision of whether or not to vaccinate is a complex one. It involves the contribution both to a social good---herd immunity---and to one's own well being. It is informed by social influence, personal experience, education, and mass media. In our work, we investigate a situation in which individuals make their choice based on how social neighbourhood responded to previous epidemics. We do this by proposing a minimalistic model using components from game theory, network theory and the modelling of epidemic spreading, and opinion dynamics. Individuals can use the information about the neighbourhood in two ways---either they follow the majority or the best-performing neighbour. Furthermore, we let individuals learn which of these two decision-making strategies to follow from their experience. Our results show that the flexibility of individuals to choose how to integrate information from the neighbourhood increases the vaccine uptake and decreases the epidemic severity if the following conditions are fulfilled. First, the initial fraction of individuals who imitate the neighbourhood majority should be limited, and second, the memory of previous outbreaks should be sufficiently long. These results have implications for the acceptance of novel vaccines and raising awareness about vaccination, while also pointing to promising future research directions.
\vspace{1.0cm}
\end{abstract}

\maketitle

\section{Introduction}
\label{sec:intro}

The development of vaccines is one of the greatest achievements of modern medicine. They save millions of lives yearly, not only by giving immunity to people exposed to an infection but also by stopping disease outbreaks. Most famously, perhaps, vaccine drove the eradication of smallpox.\textsuperscript{\cite{henderson2009smallpox}} At the time of writing, vaccines are the main hope for a pharmaceutical solution to the COVID-19 crisis.\textsuperscript{\cite{lurie,corey2020strategic, graham2020rapid}}

To be able to evaluate interventions involving vaccination, we need to model the selection of who gets vaccinated and how that affects epidemics.\textsuperscript{\cite{hufnagel2004forecast}} Creating such models is a very challenging task to which this paper seeks to contribute.\textsuperscript{\cite{WANG20151,wang2016statistical}} The main difficulty lies in the complex feedback mechanisms between the epidemics itself and the decision to get vaccinated.\textsuperscript{\cite{geoffard1997disease, chapman2012using, ibuka}} Not only are there irrational anti-vaccination sentiments that themselves spread through social contagion,\textsuperscript{\cite{jolley2014effects, saint2013vaccine}} but sometimes not getting vaccinated is a perfectly rational choice.

When the population level of immunity is high enough, an outbreak will die out by itself. If this is the case, the population is said to have \emph{herd immunity}. Even if the vaccine is effective, the marginal benefit of getting vaccinated in a society with herd immunity is small.\textsuperscript{\cite{francis1997dynamic}} On top of this, a vaccine could be costly, inefficient, laden with side-effects, or inconvenient to administer.\textsuperscript{\cite{VERELST2018106}} For an individual, the reality is often between these extremes and not a choice between a cheap lifesaver or a costly unnecessity. Thus, this choice is known in the literature as the \emph{vaccination dilemma}.\textsuperscript{\cite{fu2011imitation, cardillo2013evolutionary}}

Assume the vaccine is effective but has some side-effects (although much milder than the disease itself). Then one can model the rational-choice aspect of vaccination within the framework of game theory.\textsuperscript{\cite{wang2016statistical,chang:rev}} As often is the case, behavioural and economic game theorists both take an interest in this problem, with somewhat different starting points. The behavioural game-theory line of research typically focuses on herd immunity as a public good---something valuable and accessible without competition to anyone in society.\textsuperscript{\cite{ibuka,chapman2012using}} Like other public goods, herd immunity is prone to free-riding people who undermine the good by avoiding vaccination.\textsuperscript{\cite{coleman1988free}} Economic game theorists instead think of vaccination as a decision based on an individual's costs and benefits, and regard herd immunity as a positive externality. The truth lies in between these pictures.\textsuperscript{\cite{geoffard1997disease}} Benefiting a public good is not the most common driving force behind an individual's vaccination decisions.\textsuperscript{\cite{betsch2015using, salmon2015vaccine}} It would also make little sense to get vaccinated to contribute to herd immunity if very few others were vaccinated, which is precisely when the private benefits are the largest.\textsuperscript{\cite{francis1997dynamic}} On the other hand, herd immunity and, ultimately, eradication of a disease are the primary goals at a national level,\textsuperscript{\cite{corey2020strategic, graham2020rapid}} and thus more than a mere externality.

\begin{figure*}
\includegraphics[width=1.5\columnwidth]{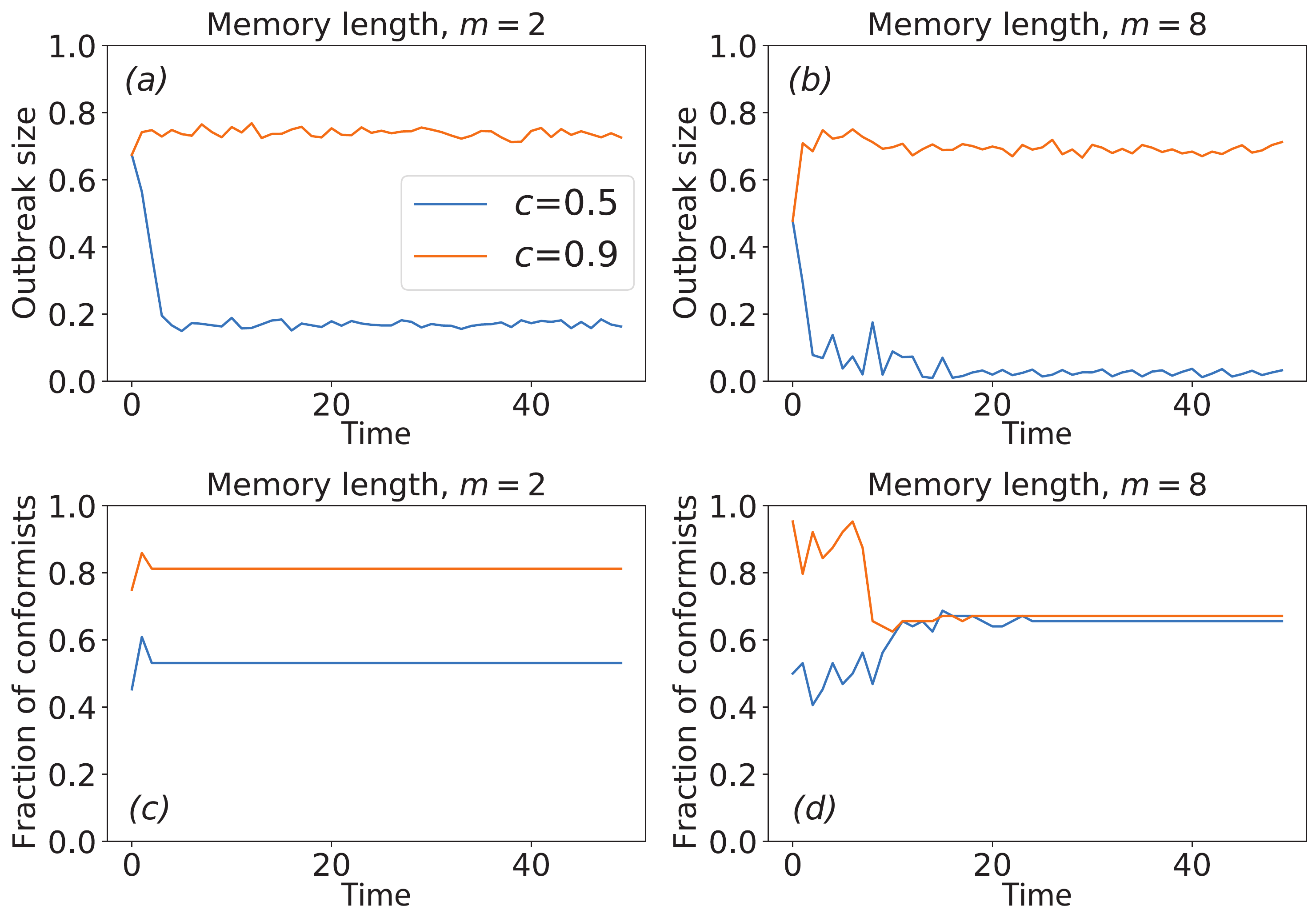}
\caption{\textbf{Time evolution of a simulation run.} We show the outbreak size ((a) and (b)) and the fraction of conformists ((c) and (d)) for a short $m=2$ ((a) and (c)) and a long $m=8$ ((b) and (d)) memory length. The plots also show two initial fractions of conformists, $c=0.5$ (blue curves) and $c=0.9$ (yellow curves). The population seems unable (respectively, able) to eventually reduce the outbreak size irrespective of the memory length when the initial fraction of conformists is high (moderate). Short (long) memory seems insufficient (sufficient) to trigger a substantial interchange of imitation mechanisms among individuals compared to the initial state. Long memory thus helps to eliminate outbreaks when the initial fraction of conformists is moderate (blue curve in panel (b)). Here, the infection rate is relatively large, $\beta=2$. The networks are constructed by the Erd\H{o}s–R\'{e}nyi model with $N=64$ nodes and the average degree $k=2$, which ensures the existence of a giant component (that essentially all nodes belong to). Time is measured in the number of vaccination cycles.}
\label{f01}
\end{figure*}

There is a growing research interest in game-theoretic studies of vaccination.\textsuperscript{\cite{chang:rev,WANG20151}} For articles in the economic game-theory literature, see e.g.\ Geoffard and Philipson\textsuperscript{\cite{geoffard1997disease}} or Francis\textsuperscript{\cite{francis1997dynamic}} and further references therein. In the evolutionary game-theory literature, early works coupled game theory and epidemic dynamics by differential-equation based models.\textsuperscript{\cite{bauch2004vaccination}} Later, authors recognised that social interaction structures are better modelled by networks. For example, in Fu et al.\textsuperscript{\cite{fu2011imitation}}\ individuals compare their fitness to randomly selected network neighbours to determine whether or not to imitate the neighbour. The phrases ``fitness'' and ``payoff'' (that in this paper are synonymous) come from the game-theory literature and capture the ability to avoid infection minus the cost associated with the vaccination. Other authors have extended the use of imitation dynamics. Zhang et al.\textsuperscript{\cite{zhang2013impact}}, for example, considered the possibility that decisions are neighbour-dependent by defining an individual's fitness depending on their entire neighbourhood. Zhang et al.\textsuperscript{\cite{zhang2012rational}}\ and Han et al.\textsuperscript{\cite{han2014can}}\ studied individuals with memory, calculating an individual's fitness as the weighted average of the past payoffs. Moreover, models by Xia et al.,\textsuperscript{\cite{xia2013computational}} Ichinose and Kurisaku,\textsuperscript{\cite{ichinose2017positive}} and Iwamura and Tanimoto\textsuperscript{\cite{iwamura2018realistic}} disregarded payoffs altogether in certain situations by allowing individuals to follow the neighbourhood majority with a non-zero probability, or, otherwise, use fitness comparison to decide their action. 

Our study extends the above models by unifying two ideas. First, when making a vaccination decision based on the performance of their neighbours, individuals can follow different decision rules. Second, an individual can learn by experience and thus change its decision rules as time goes on. We use two empirically observed decision rules. First, individuals can follow the best performer in the extended neighbourhood (the neighbourhood of a focal node and the node itself).\textsuperscript{\cite{traulsen2010human, grujic2020people}} We call such individuals \textit{performists}. Second, they can also follow the majority in the extended neighbourhood.\textsuperscript{\cite{traulsen2010human}} We will refer to individuals following this strategy as \textit{conformists}. Furthermore, we will let the individuals chose between these two rationales based on experience. The duration of the individuals' memory is one of the model-parameters we explore. In the remainder of this article, we will go over the technical details of our model, present our simulation results, and discuss the broader implications of our findings.

\begin{SCfigure*}
\includegraphics[width=1.5\columnwidth]{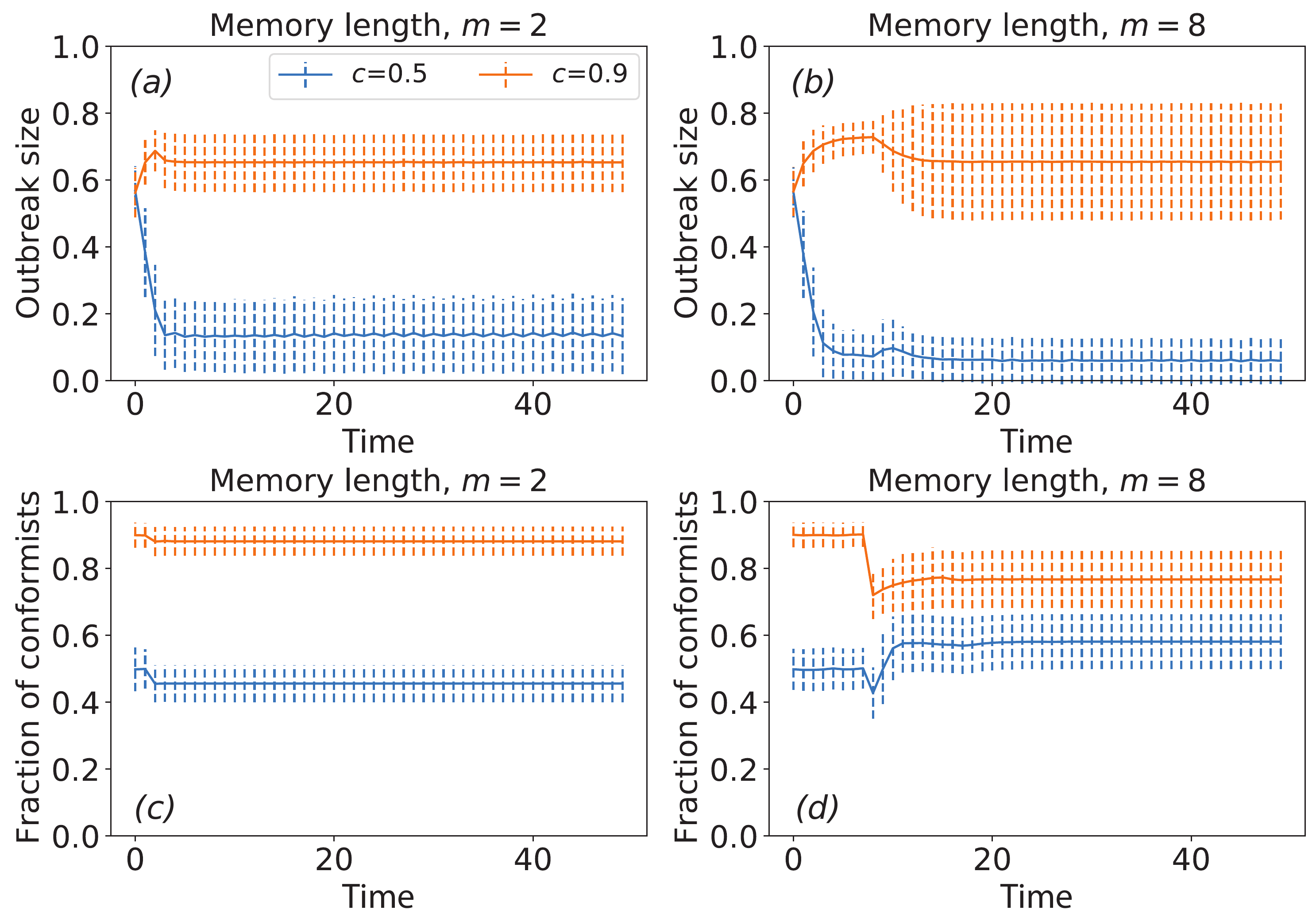}
\caption{\textbf{Time evolution of typical runs.} We show the outbreak size ((a) and (b)) and the fraction of conformists ((c) and (d)) for a short $m=2$ ((a) and (c)) and a long $m=8$ ((b) and (d)) memory length, and for two initial fractions of conformists, $c=0.5$ (blue curves) and $c=0.9$ (yellow curves). Solid curves represent ensemble averages, whereas error bars show standard deviations. The parameter values are the same as in Fig.~\ref{f01}. Time is measured in the number of vaccination cycles.}
\label{f02}
\end{SCfigure*}

\begin{figure}[!b]
\includegraphics[width=0.95\columnwidth]{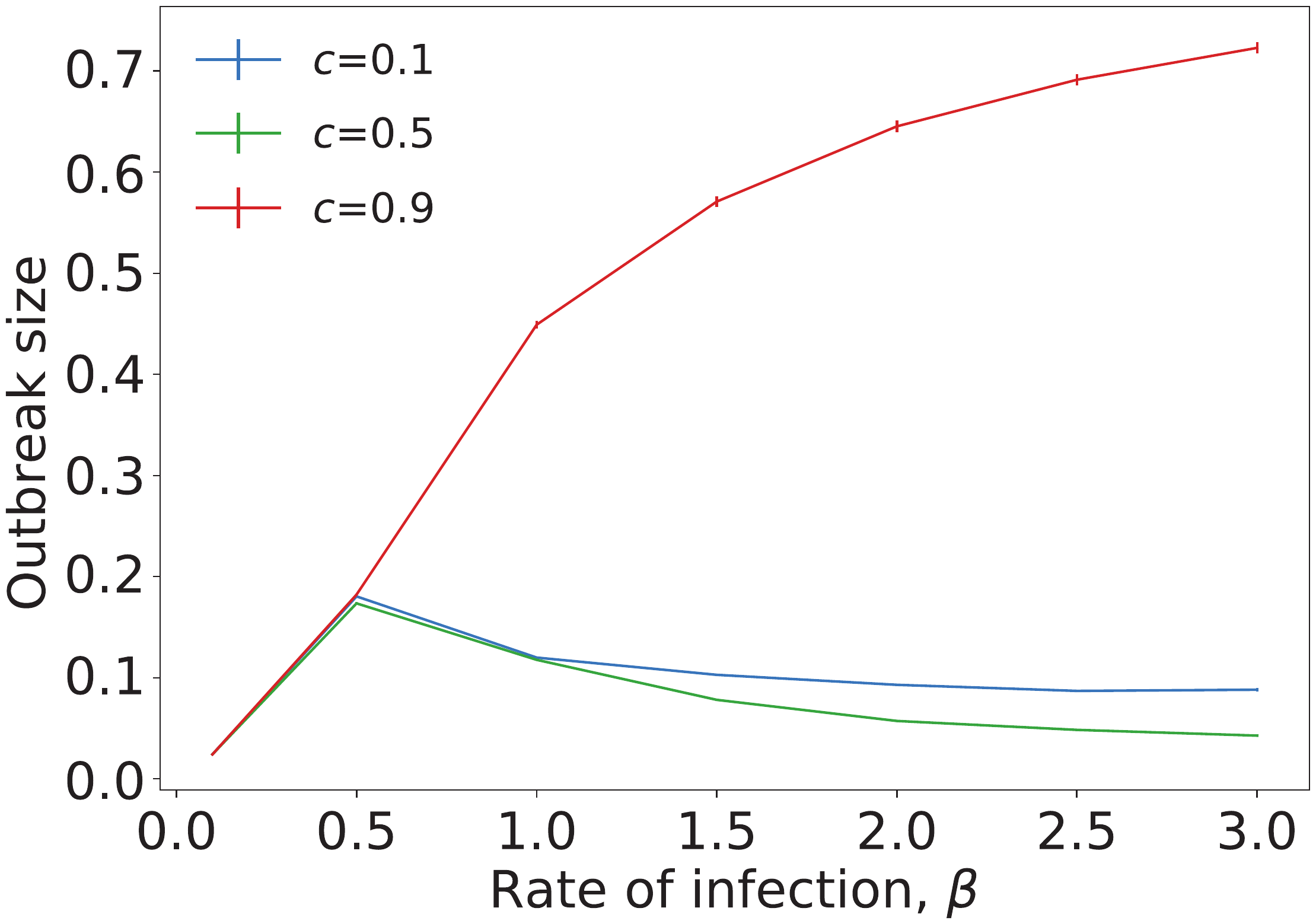}
\caption{\textbf{Low initial conformism is good, moderate initial conformism is best, and high initial conformism is bad.} We show the outbreak size, averaged over the last 50 vaccination cycles, as a function of the infection rate. The three curves correspond to three different levels of initial conformism. Low initial conformism yields good results in terms of the outbreak size, especially at large infection rates ($c=0.1$). The results are even better for moderate initial conformism, although the improvement is modest ($c=0.5$). If initial conformism is high, however, the outbreak size monotonically increases with the infection rate, and the disease may spread through much of the population ($c=0.9$). Here, the memory length is relatively long at $m=8$.}
\label{f03}
\end{figure}

\section{Model description}
\label{sec:model}

We construct a network-based epidemiological model with vaccination. The model comprises individuals embedded in a social network through which epidemics spread. Probably the closest scenario is that of seasonal influenza. Occasionally, the population has the opportunity of getting vaccinated. Then individuals evaluate the performance of their social surroundings, and, based on this information, decide about their vaccination. They have two ways (imitation mechanisms) of performing this evaluation---the mentioned conformist and performist strategies. The individuals then update these imitation mechanisms based on experience.

\begin{figure*}
\includegraphics[width=1.5\columnwidth]{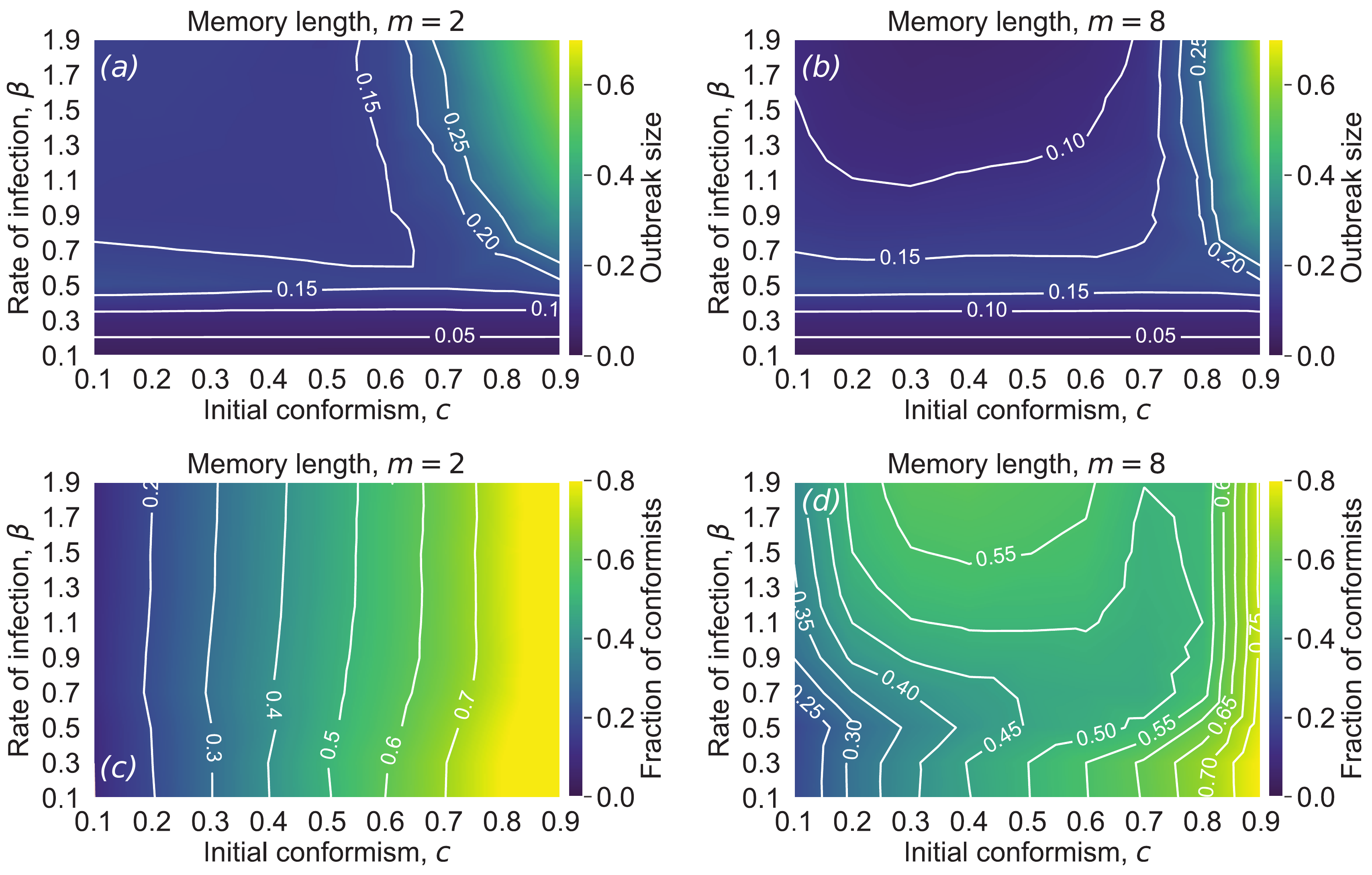}
\caption{\textbf{Updating of imitation mechanisms often leads to disease-curbing, moderate levels of conformism, but only if memory is long enough.} We show heatmaps of the outbreak size ((a) and (b)) and the fraction of conformists ((c) and (d)) as a function of the infection rate and initial conformism. When the infection rate is low, the outbreak size is independent of the initial conformism, which is evident from the horizontal contour curves for $\beta\lessapprox 0.5$ in the panels (a) and (b). As the infection rate increases, however, high initial conformism causes severe outbreaks. Short memory fails to improve the situation because the level of conformism changes little compared to initial conformism, which is seen in the almost vertical contour curves in (c). Long memory, by contrast, often improves the situation by adjusting initial conformism to disease-curbing, moderate levels between 40--60\,\%, as seen in panel (d). Every point is averaged over the last 50 vaccination cycles.}
\label{f04}
\end{figure*}

\begin{figure}[!b]
\includegraphics[width=0.95\columnwidth]{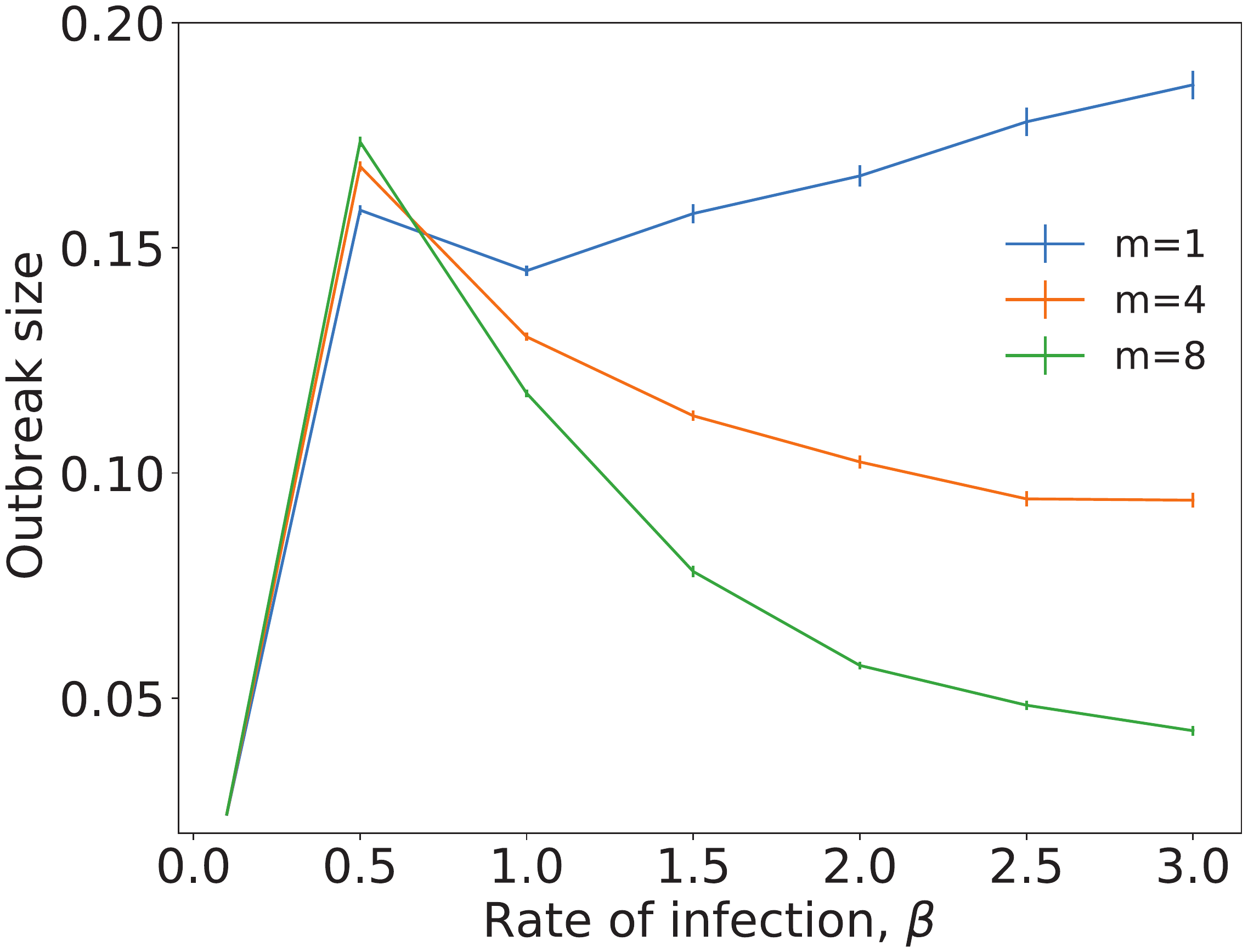}
\caption{\textbf{Long memory is better than short.} The figure shows the average outbreak size as a function of the infection rate. The three curves correspond to three different memory lengths. The outbreak size for infection rates $\beta\gtrapprox0.5$ gets progressively smaller as the memory length increases. Here, initial conformism is moderate at $c=0.5$.}
\label{f05}
\end{figure}

Each cycle of epidemic outbreaks and vaccinations plays out in three stages. First, individuals choose whether or not to vaccinate following their current strategy and available information. Second, we calculate the expectation value of the outbreak size for the particular configuration of vaccinated nodes. This is done by averaging over $640$ runs of the Susceptible-Infectious-Recovered (SIR) simulation (discussed further below). Finally, individuals choose between the conformist or performist imitation strategy.  We simulate these vaccination cycles over 150 times to reach a steady state.

At the beginning of every simulation, we randomly vaccinate  10\,\% of the population and let a fraction $c$ be conformists (otherwise performists). From the second vaccination, all individuals make decisions following their strategies. We allow the choice of strategy to depend on the experience gathered through vaccination. We assume that individuals choose the imitation mechanism that gives them a larger average payoff over the period of $m$ preceding vaccination cycles.

For the epidemic simulation, we used the standard Markovian Susceptible-Infectious-Recovered (SIR) algorithm.\textsuperscript{\cite{kiss2017mathematics}} This is the canonical model of diseases that make people immune upon recovery. It has nominally two parameters---the infection rate $\beta$ and the recovery rate $\nu$. However, $\nu$ only sets the time scale of the outbreak, and since we are only interested in the final outbreak size, we can set $\nu=1$ without loss of generality. We simulate this model with a fast ``event-driven algorithm'' as described in Kiss et al.\textsuperscript{\cite{kiss2017mathematics}} and implemented at \url{github.com/pholme/sir}.

When the individuals evaluate their performance, they assume a cost $c_\mathrm{V}$ per vaccination and a cost $c_\mathrm{I}$ for getting infected. For an approved vaccine, we can assume $c_\mathrm{V}<c_\mathrm{I}$. Accordingly, a vaccinated individual receives the payoff $\Pi_i=-c_\mathrm{V}$ per vaccination cycle, whereas someone who is not vaccinated gets the payoff $\Pi_i=-c_\mathrm{I} I(i)$, where $I(i)$ is the fraction of outbreak simulations when $i$ got infected. We set $c_\mathrm{V}=0.1$ and $c_\mathrm{I} = 1$ throughout our study.

For underlying interaction topology, we mostly use Erd\H{o}s–R\'{e}nyi random graphs with $N=64$ nodes and equally many edges. We try other, larger, sizes as well, but these give qualitatively similar results. Since this paper will not concern the large-size limit, to cut computation times, we stick with the smaller systems.

We will primarily explore three parameters in our simulations---$c$, $m$, and $\beta$ and average the simulations described above over $10^3$ simulation runs. We use the OACIS framework\textsuperscript{\cite{murase2017open}} to manage our simulations.

\begin{figure*}
\includegraphics[width=1.5\columnwidth]{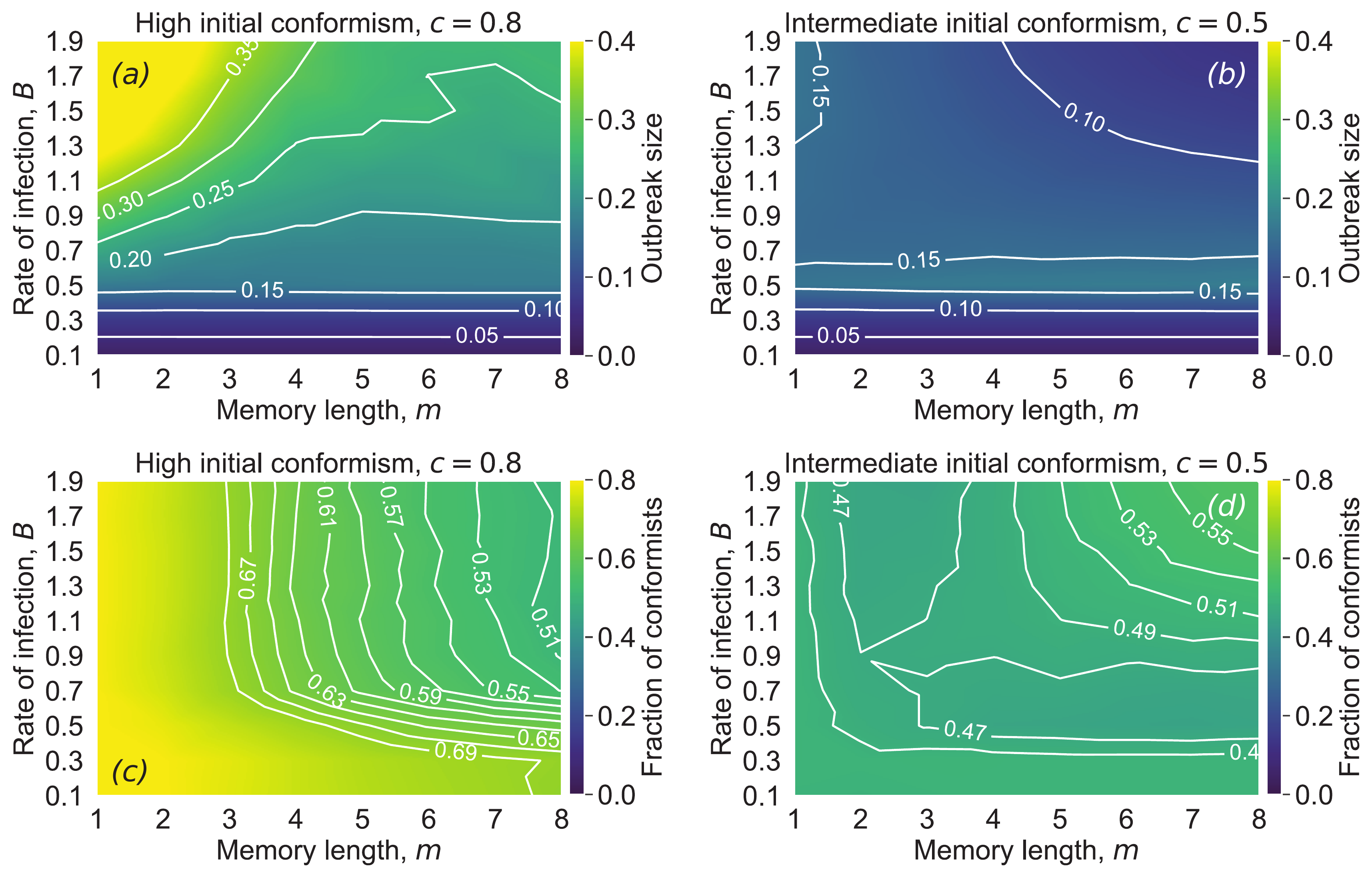}
\caption{\textbf{The level of conformism evolves only if memory is long enough and initial conformism is not moderate.} In panels (a) and (b), we show heatmaps of the outbreak size. In panels (c) and (d), we show the fraction of conformists as the functions of the infection rate and the memory length. As before, when the infection rate is low, the outbreak size is independent of the memory length, which can be seen by the horizontal contour curves for $\beta\lessapprox0.5$ in (a) and (b). As the infection rate increases, however, short memory precipitates severe outbreaks under non-moderate (a), but not under moderate initial conformism (b). The reason for this is seen in (c) in which short memory, $m<3$, fails to substantially change the level of conformism, whereas long memory, $m>4$, drives conformism to moderate levels between 40--60\,\% (for $\beta\gtrapprox0.5$). Panel (d) shows that a population starting with moderate initial conformism also ends with similar conformism levels between 40--60\,\%, irrespective of the parameters $\beta$ and $m$.}
\label{f06}
\end{figure*}

\section{Results}
\label{sec:results}

\paragraph*{Time evolution.} In Fig.~\ref{f01}, we show a representative simulation run of the model. We find that the outbreak size cannot be reduced by having a large initial fraction of conformists, irrespective of the memory length (Fig.~\ref{f01}, yellow curves in (a) and (b)). Conversely, if the initial presence of conformists is moderate, the outbreak size is greatly reduced (Fig.~\ref{f01}, blue curves in (a) and (b)). Short memory turns out to have almost no effect on the choice of imitation mechanisms, causing the population to end with a similar fraction of conformists as in the initial state (Fig.~\ref{f01}(c)). Increasing memory changes the situation. In this case, the updating of imitation mechanisms increases, allowing the population to converge to a fraction of conformists noticeably different than in the initial state (Fig.~\ref{f01}(d)). The population is even able to eliminate outbreaks in the case when the initial fraction of conformists is moderate (Fig.~\ref{f01}, blue curve in panel (b)). We confirm that the selected simulation run is indeed representative by averaging across a large number of runs (Fig.~\ref{f02}), which exhibit the same qualitative characteristics as the described single run.

\paragraph*{Initial conformism.} To further clarify the role of initial conformism, we inspect how the outbreak size depends on the infection rate for various values of the parameter $c$ when memory is relatively long (Fig.~\ref{f03}). At low infection rates, $\beta\lessapprox 0.5$, the outbreak size is independent of initial conformism. After that, for $\beta\gtrapprox 0.5$, the results diverge. High initial conformism precipitates massive outbreaks, whereas moderate ($0.4<c<0.6$) and low initial conformism prevent them. Interestingly, the outbreak size is smallest when initial conformism hovers around moderate levels of $c$.

\begin{SCfigure*}
\includegraphics[width=1.5\columnwidth]{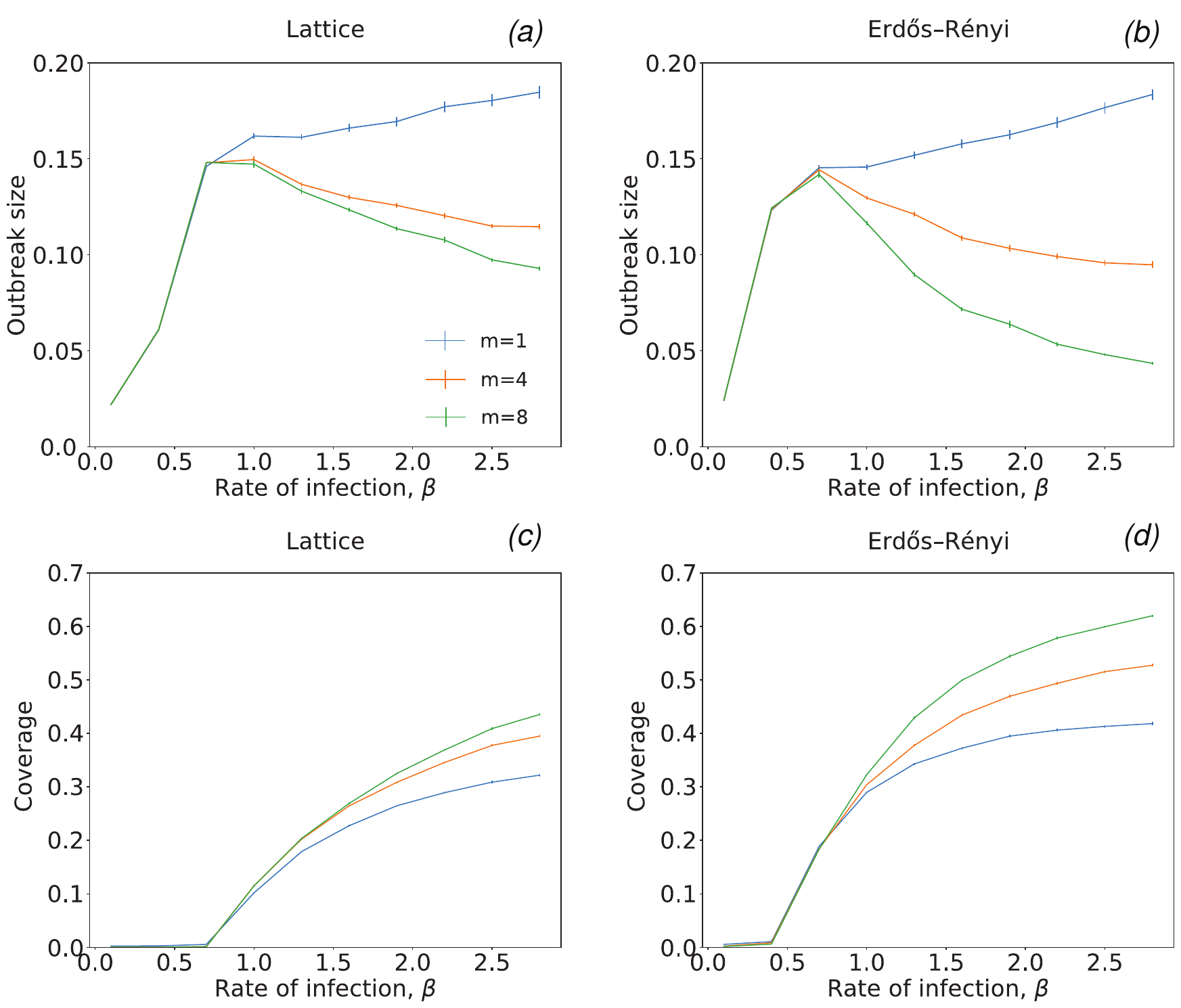}
\caption{\textbf{Network structure quantitatively impacts the results, but general conclusions stand.} In panels (a) and (b), we display the outbreak size and in panels (c) and (d) the vaccination coverage as functions of the rate of infection for two different network structures. Panels (a) and (c) pertain to the square lattice with a periodic boundary, whereas panels (b) and (d) pertain to the Erd\H{o}s–R\'{e}nyi network of the average degree $k=2$. Our observation that a longer memory length decreases the outbreak size as the rate of infection increases is valid irrespective of the network structure. Quantitative differences between networks arise because the lattice has a large diameter that impedes outbreaks given the same infection rate. Initial conformism is $c=0.50$. The number of nodes is $N = 64$.}
\label{f07}
\end{SCfigure*}

A detailed examination of the phase space reveals that a population of individuals who dynamically interchange imitation mechanisms, as envisioned in our model, often evolves to moderate levels of conformism, but only if memory is long enough (Fig.~\ref{f04}). The outbreak size is thus reduced in populations with longer memory relative to those with shorter memory, for all infection rates $\beta\gtrapprox0.5$ and irrespective of initial conformism (Fig.~\ref{f04}(a) and (b)). The reason for this reduction is that individuals who keep a shorter track of the past tend to stick to a single imitation mechanism, impeding the evolution of conformism towards moderate levels that help curb the disease (Fig.~\ref{f04}(c)). Conversely, individuals who keep a longer track of the past see their population evolve towards the disease-curbing, moderate levels of conformism between 40--60\,\% over a wide area of the phase space, that is, for $\beta\gtrapprox0.5$ and $0.10\lessapprox c\lessapprox0.85$ (Fig.~\ref{f04}(d)).

\paragraph*{Memory length.} Our results so far have suggested that memory length plays a vital role in shaping the fate of disease outbreaks. We now inspect this role in further detail. In particular, looking at how the population performs as the memory length increases, we find that long memory is better than short, even in situations when initial conformism is in the disease-curbing, moderate range (Fig.~\ref{f05}). The effect is even more pronounced when initial conformism is outside of the moderate range (Fig.~\ref{f06}(a)), the reason being that long memory, characterised by $m>4$, entices a substantial dynamical interchange of imitation rules when the infection rate is sufficiently large ($\beta\gtrapprox0.5$), thus driving the level of conformism to evolve to moderate values between 40--60\,\% (Fig.~\ref{f06}(c)). Short memory, characterised by $m<3$, cannot achieve the same as initial conformism is largely preserved throughout simulation runs (Fig.~\ref{f06}(c)). That moderate conformism indeed curbs the disease effectively is revealed by simulation runs for $c=0.5$ because in this case, the outbreak size (Fig.~\ref{f06}(b)) is always small, while the final level of conformism is still between 40--60\,\% (Fig.~\ref{f06}(d)), independent of the infection rate and the memory length.

\paragraph*{Network structure.} To confirm that the described results are robust to the choice of network structure, we have run additional simulations on two-dimensional square lattices and scale-free networks. Interestingly, only the former structure exhibits any noteworthy differences compared to the  Erd\H{o}s-R\'{e}nyi networks. Our previous observation that a longer memory length reduces the outbreak size as the infection rate increases is valid for all networks structures (Fig.~\ref{f07}(a) and (b)). The square lattice quantitatively differs from the Erd\H{o}s-R\'{e}nyi network in that the former has a much larger diameter which, for a given infection rate, impedes disease outbreaks. This, in turn, somewhat weakens the effect of long memory on suppressing diseases. Indeed, a longer memory always leads to more vaccination coverage, but such coverage is much smaller in the lattice than in the Erd\H{o}s-R\'{e}nyi network when infection rates are relatively large (Fig.~\ref{f07}(c) and (d)). Individuals thus have an easier time learning the best course of action to combat epidemics in tighter, more compact networks.

\section{Discussion}
\label{sec:disc}

We have constructed a model of vaccination---based in equal proportions on game theory, network epidemiology, and models of social influence---in which individuals in addition to deciding whether to vaccinate or not, also have the option to choose a preferred imitation mechanism. Individuals called conformists thus rely on a simple heuristic by which they imitate the behaviour of the neighbourhood majority. Performists, by contrast, imitate neighbours that perform the best in terms of payoff. The results show that the dynamic interchange of these imitation mechanisms suppresses disease outbreaks through better vaccination. Two notable phenomena are that (i) too much initial conformism in the population and (ii) short memory of individuals are obstructive for the vaccination coverage.

The reason why too much conformism leads to a low vaccination coverage is that too many individuals imitate what they see in their neighbourhoods, and they mostly see defection. This, in turn, is because our simulations start with the initial vaccination coverage of 10\,\%, which is a valid starting point from the perspective of novel vaccine acceptance. Such an angle is relevant at the time of writing. On the one hand, we are amidst a pandemic (COVID-19) for which vaccines seem like the best hope for a solution.\textsuperscript{\cite{lurie, corey2020strategic, graham2020rapid}} On the other hand, recent years have seen a trend of increasing vaccine hesitancy.\textsuperscript{\cite{dube2013vaccine, salmon2015vaccine, ibuka}} This trend could partly be explained by the kind of mix of rational thinking and predisposition to following the crowd as is manifested by the conformists.\textsuperscript{\cite{jolley2014effects, berinsky2017rumors}} However, probably it would be more appropriate to extend our model to include zealots---individuals who do not let the dynamics of the game to affect their choices.\textsuperscript{\cite{VERELST2018106, coleman1988free}}

The other notable phenomenon revealed by our model, specifically, that more extended memory helps boost the vaccination coverage, is suggestive in the sense that educational measures could be used to bolster the collective awareness of the role of vaccines in controlling infectious diseases.\textsuperscript{\cite{betsch2015using}} Classrooms are an ideal setting to raise awareness about the burden of infectious diseases as well as the success of vaccines in limiting these.\textsuperscript{\cite{henderson2009smallpox, roush2007historical}}

Finally, we note that in our simulations, the individuals learn strategies to make the vaccination choice. This means they are manifesting ``wisdom of crowds''\textsuperscript{\cite{surowiecki2005wisdom}}---i.e., that populations can perform distributed computation tasks, integrating information without central control.

\section*{Article information}

\paragraph*{Acknowledgements} Authors appreciate the support from the Japan Society for the Promotion of Science (KAKENHI grants 20H04288 and 18H01655) and the Sumitomo Foundation (grant for basic science research projects).
\paragraph*{Author contributions} M.\,J.\ and P.\,H.\ devised research. S.\,M.\ performed research. All authors discussed the results and wrote the manuscript.
\paragraph*{Conflict of interest} Authors declare no conflict of interest.

\bibliography{vaccination_game}

\end{document}